\documentclass[showpacs,aps,graphicx,twocolumn]{revtex4}
\usepackage{graphicx}
\begin{document}

\title{Economical Quantum Secure Direct Communication Network with Single Photons\footnote{Published in \emph{Chin. Phys.}
\textbf{16} (12), 3553-3559 (2007)}}

\author{Fu-Guo Deng$^{a)b)c)d)}$,
Xi-Han Li$^{a)b)}$, Chun-Yan Li$^{a)b)}$, Ping Zhou$^{a)b)}$, and
Hong-Yu Zhou$^{a)b)c)}$}
\address{$^{a)}$ The Key Laboratory of Beam Technology and Material
Modification of Ministry of Education, Beijing Normal University,
Beijing 100875,
China\\
$^{b)}$ Institute of Low Energy Nuclear Physics, and Department of
Material Science and Engineering, Beijing Normal University,
Beijing 100875, China\\
$^{c)}$ Beijing Radiation Center, Beijing 100875, China\\
$^{d}$ Department of Physics, Beijing Normal University, Beijing
100875, China}
\date{\today }

\begin{abstract}
A scheme  for quantum secure direct communication (QSDC) network  is
proposed with a sequence of polarized single photons. The single
photons are prepared originally in the same state $\vert 0\rangle$
by the servers on the network, which will reduce the difficulty for
the legitimate users to check eavesdropping largely. The users code
the information on the single photons with two unitary operations
which do not change their measuring bases. Some decoy photons, which
are produced by operating the sample photons with a Hadamard, are
used for preventing a potentially dishonest server from
eavesdropping the quantum lines freely. This scheme is an economical
one as it is the easiest way for QSDC network communication
securely.

\end{abstract}
 \maketitle

\section{introduction}

The combination of the principles in quantum mechanics and the
theory of information produces some novel applications, such as
quantum computer and quantum communication \cite{book}. Quantum key
distribution (QKD) \cite{Gisin} provides a secure way for two remote
parties, Bob and Charlie to create a private key with which they can
communicate securely by using Vernam one-time pad crypto-system. The
non-cloning theorem and the quantum correlation of an entangled
quantum system ensure the security of QKD. In 1984, Bennett and
Brassard (BB84) \cite{BB84} proposed an original point-to-point QKD
scheme based on the non-cloning theorem. As the state of each single
photon is produced by choosing randomly one of the two measuring
bases (MBs), the rectilinear basis $\sigma_z$ and the diagonal basis
$\sigma_x$, a vicious eavesdropper, Eve will inevitably disturb the
quantum systems and leave a mark in the results if she eavesdrops
the quantum line. Moreover, she cannot obtain all the information
about the single-photon state because an unknown quantum state
cannot be cloned. Now, there have been some point-to-point QKD
schemes proposed
\cite{Gisin,BB84,LongLiu,CORE,BidQKD,delay,wl02,wl03}.

Recently, a novel concept, quantum secure direct communication
(QSDC) was prosed and actively pursued by some groups
\cite{imoto1,imoto2,beige,bf,two-step,zhangzj,Wangc,wangc2,lixhcp,yan,gaot,wangj,Nguyen,report,
liconference,LIDSQC,leepra,QOTP,caisinglephoton,zhangs,lipra,dengcp}.
With QSDC, the two authorized users, say the sender Bob and the
receiver Charlie can exchange their secret message directly without
creating a private key to encrypting the message. There are two
types of QSDC schemes. One is based on entangled states, such as
those in Refs
\cite{imoto1,bf,two-step,zhangzj,Wangc,wangc2,lixhcp,yan,gaot,Nguyen,zhangs,lipra,report,liconference,LIDSQC,leepra,dengcp}.
The other is based on single photons. Typical such QSDC protocols
are the ones presented in Refs.
\cite{imoto2,beige,QOTP,caisinglephoton}. Obviously, the ones with
single photons are more convenient for the users at the aspect of
measurements than those with entangled states. The message in the
two point-to-point QSDC protocols proposed by Schimizu and Imoto
\cite{imoto2} and Beige \emph{et al.} \cite{beige} cannot be read
out until an additional classical bit is transmitted for each qubit.
We \cite{QOTP} proposed a QSDC scheme with a sequence of single
photons, and Cai \emph{et al.} \cite{caisinglephoton} introduced a
QSDC protocol with a single photon following some ideas in  Bennett
1992 QKD protocol \cite{B92}.

A general case of practical applications of QSDC requires that an
authorized user on a network can exchange the secret message
directly with another one. That is, QSDC network schemes are useful
in practical. There are some servers who provide the service for
preparing and measuring the quantum signal for the legitimate users
(the number is much lager than that of the servers), which will make
the quantum communication between the legitimate users more
convenient than those in point-to-point QSDC protocols
\cite{imoto1,imoto2,beige,bf,two-step,zhangzj,Wangc,wangc2,lixhcp,yan,gaot,wangj,Nguyen,report,liconference,
LIDSQC,leepra,QOTP,caisinglephoton,zhangs,lipra,dengcp} as it
reduces the setups for the normal users, the same as that in a
classical communication network. Although there are a few QSDC
network schemes existing based on entanglements
\cite{dengnetwork,linetwork,dengnetwork2,gaonetwork}, none with
single photons. Certainly, an ideal single-photon source is not
available for a practical application at present, but people believe
that it can be produce without difficulty in a nearly future with
the development of technology \cite{singlephotonsource}. Thus, it is
interesting, in theory, to study the model for QSDC network with
single photons.

In this paper, we introduce an economical QSDC network scheme with
single photons. The server provides the service for preparing and
measuring the quantum signals, a sequence of single photons $S$,
which will simplify the authorized users' setups for secret
communication largely. Different from QKD network protocols, the
server first sends the quantum signals to the receiver who encrypts
them with local unitary operations and then sends them to the
sender. The secret message is encoded directly on the single photons
after confirming their security. All the parties, including the
servers, agree that the initial states of the single photons are
$\vert 0\rangle$, which will reduce the difficulty for the users to
check eavesdropping. This QSDC network protocol is an optimal one.

\begin{figure}[!h]
\begin{center}
\includegraphics[width=8cm,angle=0]{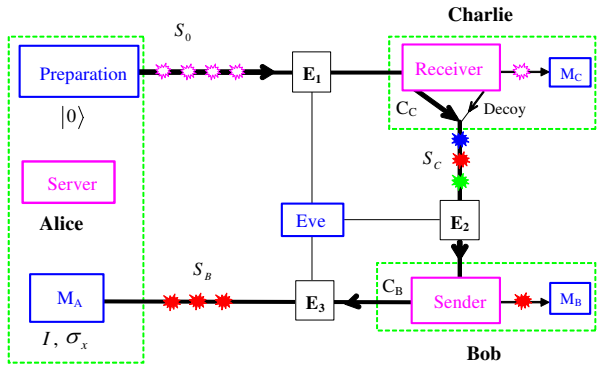} \label{f1}
\caption{ The subsystem of the present QSDC network. $M_A$, $M_B$
and $M_C$ represent the measurements done by Alice, Bob and Charlie,
respectively; $E_1$-$E_3$ represent the eavesdropping done by Eve in
different processes. $C_B$ and $C_C$ are the codes done by Bob and
Charlie, respectively. }
\end{center}
\end{figure}

\section{QSDC network with single photons}

Although a QSDC network is a composite system, its subsystem can be
simplified to three parts, the server (Alice), the sender (Bob) and
the receiver (Charlie), similar to QKD network
\cite{Phoenix,Townsend,Biham,MUQKDguo,DLMXL,LZWD}. That is, a great
number of these subsystems compose of the whole network. Similar to
the classical network communication, for each request of the
legitimate users, the server analyzes it and then performs a
relevant operation, for example, connecting the user to another one
on the network or sending a sequence of single photons to the user.
Moreover, all the users and the servers agree that the server of the
branch with the receiver provides the service for preparing and
measuring the quantum signal, and all other servers only provide the
service for connecting the quantum line for these two legitimate
users in a certain time slot if the sender and the receiver do not
exist in a branch of the network. Thus, a QSDC network scheme is
explicit if the principle of its subsystem is described clearly.

The subsystem in our QSDC network scheme is shown in Fig.1.  Alice
provides the service for preparing and measuring the polarized
single photons. The single photons $S_0$ are prepared initially in
the same states $\vert +z\rangle \equiv \vert 0\rangle$. Here $\vert
+z\rangle$ is an eigenvector of the MB $\sigma_z$. Alice sends the
photons $S_0$ first to the receiver Charlie who stores them and
checks the security of the transmission. The procedure for checking
eavesdropping includes two parts. One is used to check whether the
sample photons are in the same state $\vert 0\rangle$ or not, which
can be completed by measuring half the samples with the MB
$\sigma_z$. The other is used to check whether there are more than
one photon in each signal. That is, Charlie should forbid others
eavesdrop the quantum communication with a multi-photon fake signal
attack \cite{multiphotonattack}. This task can be accomplished with
some photon beam splitters (PBSs), similar to Ref.
\cite{multiphotonattack}. In detail, Charlie splits each signal in
the half of the samples remaining with three PBSs, shown in Fig.2,
and then measures each signal with a single-photon detector. If
there is only one photon in the initial signal, only one detector
will be clicked in theory. Otherwise, the number of the detectors
clicked is more than one with a large probability.

\begin{figure}[!h]
\begin{center}
\includegraphics[width=5cm,angle=0]{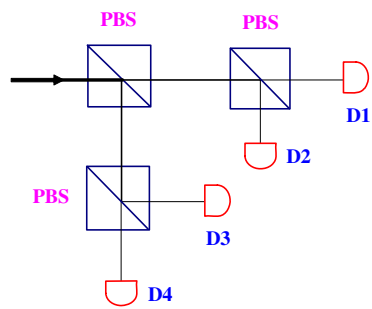} \label{f2}
\caption{ The principle of the check against multiphoton fake signal
attack, similar to Ref. \cite{multiphotonattack}. PBS presents a
photon beam splitter 50/50, and $D_i$ ($i=1,2,3,4$ ) are four
single-photon detectors. }
\end{center}
\end{figure}

If the receiver Charlie confirms that there is no eavesdropper
monitoring the quantum line between the server and him, he operates
each photon in the sequence $S_0$ with one of the two local
operations $U_0=\vert 0\rangle\langle 0\vert + \vert 1\rangle\langle
1\vert$ and $U_1=\vert 0\rangle\langle 1\vert - \vert
1\rangle\langle 0\vert$ randomly. He keeps the secret about his
operations $C_C$. In order to check eavesdropping efficiently, he
should also insert some decoy photons in the sequence $S_0$. That
is, Charlie picks up randomly some sample photons (the number is not
very large, just enough for doing the error rate analysis, about the
number of samples for eavesdropping check in BB84 QKD protocol), say
$S_{e}$ from the sequence $S_0$ after his operations with $U_0$ and
$U_1$, and performs a Hadamard ($H$) operation on each one. The $H$
operation will make the photons in $S_e$ in the state $\vert
+x\rangle=\frac{1}{\sqrt{2}}(\vert 0\rangle + \vert 1\rangle)$ or
$\vert -x\rangle=\frac{1}{\sqrt{2}}(\vert 0\rangle - \vert
1\rangle)$. That is, the states of the photons in the sequence $S_0$
are nonorthogonal, which will forbid the eavesdroppers to eavesdrop
the information on the states freely. After these operations,
Charlie sends the sequence $S_C$ (the sequence $S_0$ after Charlie's
operations) to the sender Bob.

After Bob receives the sequence $S_C$, Charlie chooses randomly some
photons from the sequence $S_C$ as the samples for checking
eavesdropping and tells their positions to Bob. The samples include
all the photons operated with a $H$ operation $S_e$ and some other
photons with only the operation $U_0$ or $U_1$. Charlie tells Bob
the MBs of the samples and requires him to announce his outcomes
obtained by measuring the samples with the same MBs as those of
Charlie's. Simultaneously, Bob can also use the same way as
Charlie's to determine whether there are more than one photons in
each sample. The error rate analysis on the samples can be
accomplished by the receiver Charlie. That is, he compares the
outcomes published by Bob with his own operations on the samples. If
they believe that the quantum line is secure, Bob codes his secret
message $C_B$ on the photons remaining in the sequence $S_C$ by
choosing the operation $U_0$ or $U_1$ according to the bit is 0 or
1, respectively. Certainly, Bob should add a small trick in the
sequence $S_C$ before he sends it out. That is, he should select
some of the photons in the sequence $S_C$ as the samples for
checking the security of the transmission between him and the server
Alice, and operates them with $U_0$ or $U_1$ randomly. Then he sends
the photons encoded, say $S_B$ to Alice who measures them with the
MB $\sigma_z$. Alice announces the outcomes of her measurements in
public, i.e., $C_A= C_C\oplus C_B$. After Bob and Charlie check
eavesdropping with the photons operated by Bob using $U_0$ and $U_1$
randomly, Charlie can read out the secret message $C_B=C_A \oplus
C_C$ directly if the transmission of whole quantum communication is
secure.

\section{security analysis}

\subsection{the relation between the information leaked and the probability detected}

In this QSDC network scheme, there are three processes for checking
eavesdropping, i.e., one for the transmission between the server
Alice and the receiver Charlie, one between Charlie and the sender
Bob, and the other one between Bob and Alice. Although an
eavesdropper, say Eve (including a dishonest server Alice), can
eavesdrop the quantum communication between the two parties of the
transmission in each process, say $E_1$, $E_2$ and $E_3$ (see Fig.
1), the eavesdropping checks can forbid her to monitor the quantum
line freely, as the process of each transmission is similar to that
in BB84 QKD protocol \cite{BB84} which is proven unconditionally
secure with error correction and privacy amplification
\cite{BB84proof}. That is, Eve's action will be detected by Charlie
and Bob before Bob encodes his message on the photons.

In detail, with the eavesdropping $E_1$, Eve can only obtain the
information about the initial states of the photons, which is known
to every one, not a secret. If Eve wants to steal the information
about the operations $C_C$, she should insert some Trojan-horse
photons in the original signal or intercept the photons operated by
Charlie and measure them. Obviously, her actions will inevitably
leave a trace in the first or second process for eavesdropping
check. The reason is that the Trojan-horse photons will be found out
when Charlie measures the samples with some PBSs as there are more
than one detector clicked for each original signal received from
Alice. If Eve exploits the intercept-resending attack to steal the
information about Charlie's operations $C_C$, her action will
introduce some errors in the outcomes of the measurements on the
samples in the second process for eavesdropping check, the same as
that in BB84 protocol \cite{BB84}. The eavesdropping on the last
stage $E_3$ (the transmission between Bob and Alice) will give Eve
nothing about the message $C_B$ as it is encrypted by the operations
$C_C$ \cite{Gisin,QOTP}. In a word, Eve cannot steal the information
about the secret message $C_B$ freely in an ideal condition. If the
noise in the quantum line is not very small, the case is different.
We discuss it as follows.

Similar to Ref. \cite{dengnetwork,LM}, we can also use the relation
of the  error rate and the correspondent maximal amount of
information obtainable for an eavesdropper from a photon with $E_2$
to demonstrate the security of the quantum channel between Charlie
and Bob. As Bob checks eavesdropping by choosing the two MBs
$\sigma_z$ and $\sigma_x$ for the samples, the same as the BB84 QKD
protocol \cite{BB84}, the eavesdropping done by Eve can be realized
by a unitary operation, say, $\hat{E}$ on a larger Hilbert space.
That is, Eve can perform the unitary transformation $\hat{E}$ on the
photon sent by Charlie (say photon $C$) and the ancilla whose state
is initially in $\vert 0\rangle$ \cite{dengnetwork,LM}.
\begin{eqnarray}
\hat{E}\vert 0\rangle_{C} \vert 0\rangle &=& \sqrt{F}\vert 0\rangle
\vert e_{00}\rangle + \sqrt{D}\vert 1\rangle \vert
e_{01}\rangle, \label{ut3}\\
\hat{E}\vert 1\rangle_{C } \vert 0\rangle &=& \sqrt{D}\vert 0\rangle
\vert e_{10}\rangle + \sqrt{F}\vert 1\rangle \vert e_{11}\rangle,\\
\hat{E}\vert +x\rangle_{C} \vert 0\rangle &=&
\frac{1}{\sqrt{2}}[\sqrt{F}(\vert 0\rangle \vert e_{00}\rangle +
\vert 1\rangle \vert e_{11}\rangle)\nonumber\\ && + \sqrt{D}(\vert
1\rangle
\vert e_{01}\rangle + \vert 0\rangle \vert e_{10}\rangle)],\\
\hat{E}\vert -x\rangle_{C} \vert 0\rangle &=&
\frac{1}{\sqrt{2}}[\sqrt{F}(\vert 0\rangle \vert e_{00}\rangle +
\vert 1\rangle \vert e_{11}\rangle)\nonumber\\ && - \sqrt{D}(\vert
1\rangle \vert e_{01}\rangle + \vert 0\rangle \vert e_{10}\rangle)],
\end{eqnarray}
where $F$ is the fidelity of the state of the photon $C$ after the
eavesdropping, $D$ is the probability that Bob and Charlie can
detect Eve's action. As  the operation $\hat{E}$ is a unitary one,
we have the relations as follows
\begin{eqnarray}
\langle e_{00} \vert e_{00} \rangle + \langle e_{01} \vert e_{01}
\rangle =  F + D =1,
\\
\langle e_{10} \vert e_{10} \rangle + \langle
e_{11} \vert e_{11} \rangle =  D  + F =1,\\
\langle e_{00} \vert e_{10} \rangle + \langle e_{01} \vert e_{11}
\rangle=0.
\end{eqnarray}

After the eavesdropping check, Bob encodes his message on the
photons in the eigenvectors of the MB $\sigma_z$ with the operations
$U_0$ and $U_1$.  As the states of the samples are random for any
eavesdropper, the photon $C$ is in the state
$\rho_{C}=\frac{1}{2}\left(
\begin{array}{cc}
1 & 0 \\
0 & 1%
\end{array}%
\right)$. After the unitary operation $\hat{E}$, the final state of
the photon $B$ and the ancilla $e$ is described as follows
\cite{dengnetwork,D2}.
\begin{eqnarray}
\varepsilon(\rho_{C})= P_0\varepsilon_{U_0}(\rho_{C}) +
P_1\varepsilon_{U_1}(\rho_{C}),
\end{eqnarray}
where $P_0$ and $P_1$ are the probabilities encoded with the
operation $U_0$ and $U_1$, and $\varepsilon_{U_0}$ and
$\varepsilon_{U_1}$ are quantum operations describing the evolution
of the initial state $\rho_{C}$, i.e.,
\begin{eqnarray}
\varepsilon_{U_i}(\rho_{C})=U_i\hat{E}\rho_{C}\otimes\vert
e\rangle\langle e\vert \hat{E}^+U^+_i.
\end{eqnarray}
The accessible information extracted from the state
$\varepsilon(\rho_{C})$ is no more than the Holevo bound
\cite{dengnetwork,D2,bf}, i.e.,
\begin{eqnarray}
I_{Eve}&\leq & S(\varepsilon(\rho_{C}))- [P_0
S(\varepsilon_{U_0}(\rho_{C})) + P_1
S(\varepsilon_{U_1}(\rho_{C}))]\nonumber\\
& \equiv & I_{max},
\end{eqnarray}
where $S(\rho)$ is the von-Neumann entropy of the state $\rho$,
i.e.,
\begin{equation}
S(\rho)=-Tr\{\rho log_2 \rho\}=\sum\limits_{i=0}^{3}-\lambda
_{i}\log _{2}\lambda _{i}, \label{information1}
\end{equation}
where $\lambda_i$ are the roots of the characteristic polynomial
det$(\rho-\lambda I)$ \cite{bf}.

Suppose that Bob chooses the same probability for choosing the two
operations $U_0$ and $U_1$, i.e., $P_0=P_1=1/2$, we can obtain the
relation between $I_{max}$ and the probability $D$ as follows when
Eve measures the composite system with the basis $\{\vert 0\rangle
\vert e_{00}\rangle$, $\vert 0\rangle \vert e_{01}\rangle$, $\vert
1\rangle \vert e_{00}\rangle$, $\vert 1\rangle  \vert e_{01}\rangle
\}$ which spans the generic subspace of the Hilbert space $H_{C}
\otimes H_e$ support of $\varepsilon(\hat{\rho}_{C})$
\begin{eqnarray}
I_{max} &=& -Dlog_2 D - (1-D)log_2(1-D).
\end{eqnarray}
That is, the probability of being detected $D$ will increase if Eve
wants to steal more information about the secret message. If the
error rate introduced by noise is not zero, Bob should exploit the
single-photon quantum privacy amplification method \cite{SPQAP} to
reduce the leakage largely.

\subsection{quantum privacy amplification for single qubits}

In brief, the principle of the quantum privacy amplification for two
single photons can be described as follows \cite{SPQAP}. It includes
two controlled-not (CNOT) gates and an $H$ gate, shown in Fig.3.
Suppose the original states of the photon 1 and photon 2 are $\vert
\varphi \rangle _{1}=a_{1}\vert 0\rangle + b_{1}\vert 1\rangle$ and
$\vert \varphi \rangle _{2}=a_{2}\vert 0\rangle +b_{2}\vert
1\rangle$, respectively. After the operation, the qubit 2 is
measured with the MB $\sigma_z$ and the information of its original
state is incorporated into photon 1 as
\begin{eqnarray}
\left\vert \psi \right\rangle _{out} &=&\frac{1}{\sqrt{2}}%
\{(a_{1}a_{2}+b_{1}b_{2})\left\vert 0\right\rangle
_{1}+(a_{1}b_{2}-b_{1}a_{2})\left\vert 1\right\rangle
_{1}\}\left\vert
0\right\rangle _{2}  \nonumber \\
&+&\frac{1}{\sqrt{2}}\{(a_{1}a_{2}-b_{1}b_{2})\left\vert
1\right\rangle _{1}+(a_{1}b_{2}+b_{1}a_{2})\left\vert
0\right\rangle _{1}\}\left\vert 1\right\rangle _{2}.\nonumber\\
\label{s3}
\end{eqnarray}
This means that the state of the control qubit $\varphi\rangle_1$
will contain the information of the state of the original target
qubit no matter what the result obtained by Bob is. In this way, the
probability that Eve gets the complete information about the output
state is reduced to $r^2$. Here $r$ is the original probability that
Eve obtains the information about the qubits in a noisy channel.

\begin{figure}[!h]
\begin{center}
\includegraphics[width=6cm,angle=0]{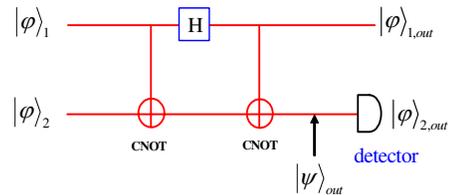} \label{f3}
\caption{ Quantum privacy amplification operation for two qubits
\cite{SPQAP}. }
\end{center}
\end{figure}

If the noise in the quantum line somewhat depolarizes the single
photons, the legitimate users can purify the polarized single
photons \cite{singlepurification} and then perform quantum privacy
amplification on them for eliminating the information leaked to Eve
\cite{SPQAP}. These two operations will eliminate the effect of the
noise. In this way, the security of this QSDC network scheme is as
the same as that with an ideal quantum line. Of course, if the noise
or the loss in the quantum line is not small, this scheme is not
good for transmitting the secret message directly, but for creating
a private key efficiently.

\section{discussion and summary}

It is of interest to point out the advantage that the sequence $S_0$
is first sent to the receiver Charlie, not the sender Bob. As the
secret message $C_B$ cannot be discarded, different from the
outcomes in QKD, the users have to confirm whether the quantum line
between them is secure or not \cite{two-step,QOTP}. Only when the
quantum line is secure, the sender Bob would encode his message on
the quantum information carriers, a sequence of polarized single
photons. The operations done by Charlie $C_C$ carry nothing about
the message before Bob codes the photons. That is, $C_C$ is just a
raw key, same as that in QKD before Bob and Charlie confirm the
security of the transmission between them, and can be discarded. But
after the confirmation is done by Bob and Charlie, the operations
$C_C$ become the unique private key for decrypting the message
$C_B$. This order of the transmissions ensures that the message
$C_B$ are not revealed to Eve even though she monitors the quantum
lines. Moreover, the two users, in this time, can exploit quantum
privacy amplification to reduce the leakage to a low level as the
photons do not carry the secret message.

Another character of this QSDC network scheme is that the initial
states prepared by the server Alice are all $\vert 0\rangle$. It
will reduce the difficulty for the parties to check eavesdropping.
In detail, Bob and Charlie can complete their eavesdropping check
without the help of Alice's, which can prevent Alice from attacking
the communication with a fake signal and cheating \cite{QSSattack}.
Moreover, Charlie can use an $H$ operation to change a photon into a
decoy one efficiently.

Compared with the QSDC network schemes existing
\cite{dengnetwork,linetwork,gaonetwork}, a sequence of polarized
single photons is enough, not entanglements. The users on the
network need only have the capability of performing single-photon
measurement and local unitary operations, not multipartite joint
measurements. Moreover, it is unnecessary for the users to have an
ideal single-photon source as the quantum signals are prepared by
the server, which will simplify the devices of the users on the
network. Same as the point-to-point QSDC scheme \cite{QOTP}, almost
all the photons can be used to carry the useful information in
theory. That is, the efficiency for qubits $\eta_q$ approaches
100\%. Except for checking eavesdropping, it is unnecessary for the
users to exchange the classical information. Thus this QSDC network
scheme is an economical one.

Certainly, on the one hand, if the sender Bob has a perfect device
for measuring a sequence of single photons, he can send no photon to
the server Alice and he measures them himself in this network
scheme. In this way, the server Alice has more space for dealing
with the request of other users'. In essence, the server Alice only
provides the service for preparing a sequence of single photons in
this modified network scheme. On the other hand, if a perfect device
for measuring single photons is very expensive in a practical
application, the sender Bob can send the photon sequence to the
sever Alice for determining their states. This structure will reduce
the requirements on the users' devices largely at the expense of
Alice's resource.

In summary, we have presented a new QSDC network scheme with a
sequence of polarized single photons $S$. In this scheme, the server
prepares all the quantum signals initially in the same state $\vert
0\rangle$, which will reduce the difficulty for the parties to check
eavesdropping. After confirming the security of the single photons
encrypted by the receiver, the sender encodes his secret message
directly on them. After the server announces the combined operations
on the photons, the receiver can read out the message directly. This
scheme is an economical one as it is the easiest way for QSDC
network communication securely. If the noise or the loss in the
quantum line is not small, the parties can use this model for
creating a private key efficiently, not exchanging the secret
message.

\section{acknowledgement}

This work is supported
 by the National Natural Science Foundation of
China (Grant Nos. 10604008 and 10435020) and the Beijing Education
Committee (Grant No XK100270454).

\end{document}